# Emergent Viscosity an Alternative for Dark Matter in Galaxies

M. B. Altaie
Department of Physics, Yarmouk University, 21163 Irbid, Jordan.
N. R. Suleiman
Department of Astronomy, Eötvös University, 1117 Budapest, Hungary

**Abstract**: We assume that the individual stars which are located at the peripheral parts of the spiral galaxies are experiencing a drag force acting upon them radially. Such a force might be produced by some sort of a dynamically generated viscous medium, and as the stars are in the state of free fall toward the center of the galaxy, then it would balance the centripetal force acting on the star, thus resulting in a terminal velocity. We make no attempt to explain the origin of the assumed drag force or show how it could be generated, but we have tried to test such an assumption by fitting the calculated velocity curves of 18 spiral galaxies with rotation curves obtained from actual observations. Results show remarkable agreements.

# Introduction

The problem of the rotation curves of a spiral galaxies is well-known since long time. Classical reviews which explains the origin of the problem since the work of Zwicky [1] are available [2]. While one would expect the stars and gases in the far parts of the galaxy to behave like being under the effect of the central field of force of the galaxy, thus following Kepler's third law, it is observed that the velocity of these parts of the galaxy is nearly constant and nearly do not change with the distance except for small variations in most cases [3], [4], and [5]. Recently, new studies have extended the range of observation to go up to 200 kpc for our galaxy [6]. Other very recent studies are giving more accurate results for the velocity profile [7] and [8] confirming the nearly constant, velocity profile for the stars far from the galactic bulge. This means that the velocities of the stars and gases in this part of the galaxy are lower than what would be expected for a solid disk and is higher than what would be expected from Kepler's third law. This might indicate that the mass of the galaxy is much more than the mass calculated from the observed matter.

The dominant explanation which was given to this problem was the assumption of the presence of invisible matter within the galaxies and clusters of galaxies. This excess mass was called *dark matter* and was thought to be within the galactic halo. Dark matter is thought to interact only through its gravitational effect. However, for this assumption to work most of the dark matter has to be located in the galactic halo. If dark matter is to be clumped in the form of dense planet-sized objects then one might assume that such object are in stable orbits around the galaxy, but the recent observations of the gravitation lensing shows that this is not the case [9]. According to the observations from the gravitational lensing, dark matter is almost homogeneously distributed within the galactic halo. Several suggestions for other the possible candidates of dark matter were presented during the last decades but the preferred one was the so-called weakly interacting massive particles (WIMPs) [10]. For this reason several projects were launched in an endeavor to detect such WIMPs (for detailed presentations see [11] and [12]). The last and the most sensitive detector of these was the project of Large Underground Xenon detector (LUX) [13]. Despite the pains taking efforts the LUX project failed to detect any signal which can be taken confidently to be due to WIMP [14]. The presence of dark matter got support from the analysis of cosmic microwave background radiation where it was found that the mass density of the universe is larger than the mass density of the observed baryonic matter [15] and [16] (see also the measurements of the Planck project [17] and results of earlier projects e.g. [18]).

Another proposal to resolve the problem of the motion of galaxies suggested a modified Newtonian dynamics (MOND) which is claimed to apply in cases of very low acceleration [19]. This is a well-studied proposal [20] [21] [22] [23] despite the fact that it lacks rigorous theoretical foundations.

The emergent gravity proposal suggests the presence of entropic force showing elastic effect that causes higher gravity than expected on bases of standard general relativity [24], So such the higher gravity causing the outer parts of the galaxy attain higher velocity is not necessarily caused by the presence of higher mass but other effects in this proposal. However a recent study has shown that the calculations based on emergent gravity do not explain the rotation curves of the galaxies [25].

Studies concerning the possible interaction of quantum field with baryonic matter are available. Away from any conclusion in this respect assume the existence of an emergent viscous force that might result from the interaction of baryonic matter of the moving stars with the virtual quantum states surrounding such stars during their motion. Such emergent force will cause a drag that will cause the star to have ultimate constant radial velocity.

## The Model

In this article we are going to test a proposal which may suggest a new model to explain the rotation curves of spiral galaxies. We assume that the individual stars which are located at the peripheral parts of the galaxy are experiencing a drag force acting upon them radially. Such a force might be produced by some sort of a dynamically generated viscous medium, and as the stars are in the state of free fall toward the center of the galaxy, then it would balance the centripetal force acting on the star, thus resulting in a terminal velocity. We will not make any attempt here to explain the origin of the assumed drag force or show how it could be generated, but will only try to test such an assumption by fitting the calculated velocity curves of some galaxies to actual observations and see if it is comply with the assumed dynamics. If the fitting results are satisfactory then the idea might become worthwhile to be considered for further studies in a more profound theoretical context.

As the galaxy rotates, its central part will certainly behave like a disk because of the high density of celestial objects in that region. The outer parts of the galaxy are in motion like any planetary system; it can be approximated to be in the state of free fall as it is moving under the acceleration of the gravity. As such is the motion of the stars, then once a viscous medium of any sort is assumed to exist, a radial drag force will be generated which will have the mechanical status similar to what happens when a metal ball is dropped into a vessel filled with oil. The viscous drag force will eventually balance the gravitational force and consequently the falling body will attain terminal velocity. In this model stars falling under the act of gravity will also attain such a terminal velocity and move with constant speed throughout its path. This proposal may solve the problem of dark matter on the cosmological scale too since the expansion of the space between large structures in the universe would generate similar effect to that taking place by motion of individual stars of the galaxy.

## Calculations and Results

We assume the presence of a drag force acting radially and balancing the gravitational force acting on the stars. Therefore, we may equate the emergent viscous force taken here to be described by Stoke's formula with the gravitational force causing the motion of the star. So, we have

$$F_d = F_g, \qquad (1)$$

this means

$$6\pi a \eta v = \frac{GmM}{r^2}, \qquad (2)$$

where $m$ is the mass of the star, $M$ is the mass of the inner part of the galaxy, $\eta$ is the coefficient of the emergent vacuum viscosity, $a$ is the radius of the star, $r$ is the distance of the star measured from the galactic center, and v is its observed velocity.

Eq. (2) may be written as

$$6\pi \left(\frac{a}{m}\right) \eta v = \frac{GM}{r^2}.$$

Assuming a circular orbit, the velocity of the star is

$$v^2 = \frac{GM}{r}, \quad (3)$$

Thus substituting (3) into (2) the velocity of the star under this mechanism of the drag force will be given by

$$v = 6\pi \left(\frac{a}{m}\right) \eta r. \quad (4)$$

For this expression to give a constant velocity for the stars in their orbits around the galactic center the viscosity coefficient $\eta$ should be proportional to $1/r$. Let us, for the sake of argument, assume a more generalized form of the variation of $\eta$ with the distance $r$ and set

$$\eta = \frac{C}{B+r} \quad (5)$$

Where $B$ and $C$ are constants that would be determined, in this model, from fitting the observations. Accordingly, the velocity of a star located at a distance $r$ from the center of the galaxy will be given by

$$v = 6\pi \left(\frac{a}{m}\right) \frac{Cr}{B+r} \quad (6)$$

We will try to test this formula by correlating it with actual observations from our galaxy, the Milky Way. In the next step, we will check if this formula fits well with observational results obtained for the rotation curves of other spiral galaxies.

Primarily we have no idea about the value of the viscosity coefficient $\eta$. So, let us first evaluate this coefficient empirically using Eq. (4) and the available data about the sun's kinematics given in [26]. The aim is to know the order of magnitude of $\eta$ in order to obtain an estimation for the viscosity of the medium near the sun. Consequently we will be able to find the viscosity function along whole Galaxy. To do this we plot the observed circular velocities $v_c$ of the stars belonging to our galaxy versus their distance from the galactic center. The basic data for the sun used in our calculations are the same as those used by ref. [26], these are: the mass $M = 1.9889 \times 10^{30}$ kg, the radius $a = 6.953 \times 10^8$ m, distance from galactic center $r = 2.57 \times 10^{20}$ m, $v = 2.5 \times 10^5$ m/s. From these basic data we can have a rough estimate of the order of magnitude of the viscosity coefficient $\eta$, using Eq.(4) we get

$$\eta = 1.47 \times 10^5 \text{ kg/m.s} \quad (7)$$

Now, in order to find an estimate of the viscosity function along the whole galaxy we plot the observed circular velocities of the stars versus its distances, this we obtain from the observational data given in [6] shown in table (1).

**Table (1)**

| $r$ (kpc) | $v_c$ (km/s) |
|---|---|
| 1.61 | 217.83 |
| 2.57 | 229.58 |
| 3.59 | 223.11 |
| 4.51 | 247.88 |
| 5.53 | 253.14 |
| 6.50 | 270.95 |
| 7.56 | 267.80 |
| 8.34 | 270.52 |
| 9.45 | 235.58 |
| 10.50 | 249.72 |
| 11.44 | 261.96 |
| 12.51 | 284.30 |
| 13.53 | 271.54 |
| 14.59 | 251.43 |
| 16.05 | 320.70 |
| 18.64 | 286.46 |

The results for the rotation curve of our galaxy is shown in Fig. (1). This distribution of velocities can be taken as a base for a simple fitting out of which we obtain a functional description of the velocity profile. Once we obtain this functional profile we will have an estimate for a trial function that might be used to check the adherence of the velocity profile of other galaxies to our model.

From the fitting of the rotation curve of our galaxy shown in Fig. (1) we obtain the circular velocity as a function of the radial distance from the galactic center. This is given by

$$v_c = \frac{286.41148}{0.68598 + r}. \qquad (8)$$

Then, using (q6) the dependence of the viscosity coefficient on the radial distance will be given by

$$\eta(r) = \frac{m}{6\pi a} \frac{286.41148}{0.68598 + r}. \qquad (9)$$

To simplify the model we will take the ratio $m/a$ for the stars to be equal that of the sun which is $2.86 \times 10^{21}$ kg/m. Accordingly, we get

$$\eta(r) = \frac{4.346 \times 10^{25}}{2.116 \times 10^{16} + r}, \qquad (10)$$

where now $r$ is in kilometers and $\eta$ is in kg/km.s.

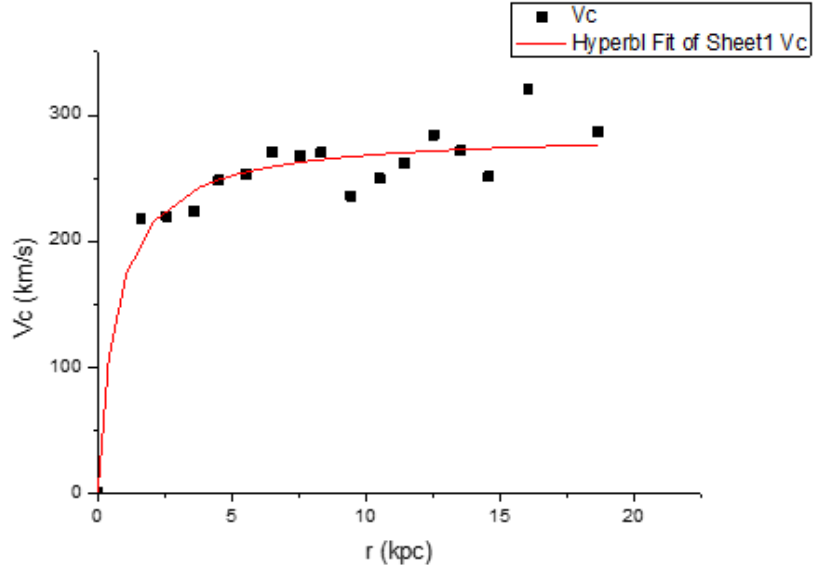

**Figure (1) Fitting of the rotation curve of Milky Way**

## Fitting the rotation curve of other galaxies

McGaugh, Lelli and Schombert [8] employ the new Spitzer Photometry and Accurate Rotation Curves (SPARC) database. SPARC is a sample of 175 disk galaxies representing all rotationally supported morphological types. It includes near-infrared observations that trace the distribution of stellar mass and 21-cm observations that trace the atomic gas. The 21-cm data also provide velocity fields from which the rotation curves are derived. In some cases, these are supplemented by high spatial resolution observations of ionized interstellar gas. SPARC is the largest galaxy sample to date with spatially resolved data on the distribution of both stars and gas as well as rotation curves for every galaxy.

We have chosen data for 18 spiral galaxies from the SPARC with different distances as shown in table (2). These galaxies are chosen from the available set with radii near that of our galaxy. We calculate the viscosity function $\eta(r)$, and then calculate the velocity function for these galaxies. For this purpose we find that the best fitting function can be expressed in terms of an exponential function as

$$v(r) = b(1 - e^{-r/c}), \qquad (11)$$

where $b$ and $c$ are constants which varies from one galaxy to another. Accordingly the viscosity coefficient function will be given by

$$\eta(r) = \frac{1}{6\pi}\left(\frac{m}{a}\right)\frac{b(1-e^{-r/c})}{r}. \qquad (12)$$

The constants $b$ and $c$ for the 18 galaxies under our consideration are given in the last two columns of table (2). It is remarkable that the constant $b$ and $c$ in (11) and (12) correspond to the terminal velocity at the far rim and the galactic bulge respectively.

**Table (2)**

| Galaxy Name | Distance (Mpc) | $b$ (km/s) | $c$ (kpc) |
|---|---|---|---|
| ESO 563 | 60.8 | 312.7 | 4.85 |
| F-568-3 | 84.4 | 122.4 | 4.86 |
| F-583-1 | 35.4 | 86.3 | 3.37 |
| IC 4202 | 100.4 | 247.1 | 3.02 |
| NGC 1090 | 37.0 | 160.0 | 2.05 |
| NGC 2903 | 6.6 | 180.6 | 0.73 |
| NGC 2998 | 68.1 | 203.0 | 1.20 |
| NGC 3198 | 13.8 | 149.0 | 2.58 |
| NGC 4559 | 9.0 | 119.1 | 2.07 |
| NGC 6015 | 17.0 | 152.0 | 1.27 |
| NGC 6503 | 6.2 | 115.0 | 0.70 |
| UGC 00128 | 64.5 | 125.2 | 4.19 |
| UGC 01230 | 53.7 | 103.3 | 22.2 |
| UGC 03205 | 50.0 | 220.0 | 0.91 |
| UGC 03580 | 20.7 | 124.4 | 1.69 |
| UGC 06786 | 29.3 | 211.5 | 0.23 |
| UGC 11455 | 78.6 | 266.0 | 4.03 |
| UGC 12506 | 100.6 | 225.0 | 2.10 |
| Milky Way | 0 | 266.9 | 1.9 |

Below we plot the calculated fitting curves against the observational data for the 18 galaxies listed in table (2) where the square dots stand for the observational data and the fitting is represented by the solid line. It is interesting to note that some curves shows excellent fit even better than that obtained for our galaxy shown in Fig. (1). The remarkable fact of our work is the ability to adopt the fitting of the rotation curves of so many galaxies with the identification of two parameters only, the radius of the galactic bulge and the terminal velocity of their far rims.

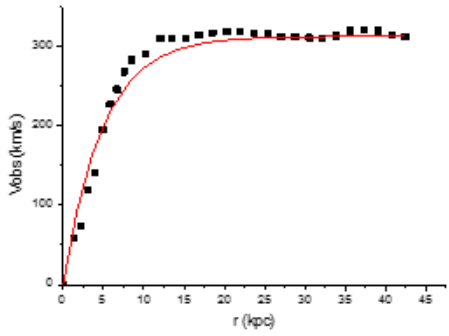
Fig. (2) ESO 563

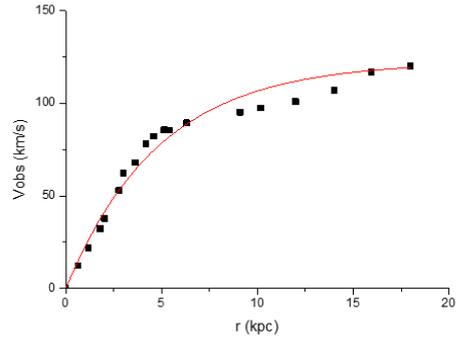
Fig. (3) F586-3

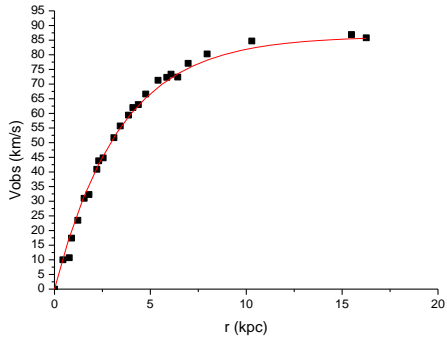
Fig. (4) F 583-1

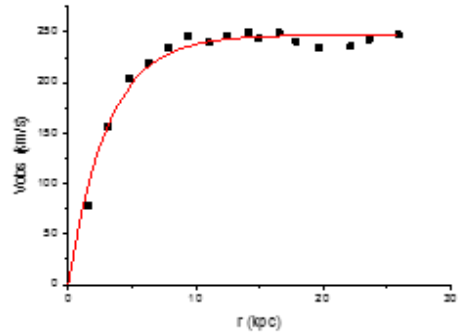
Fig. (5) IC 4202

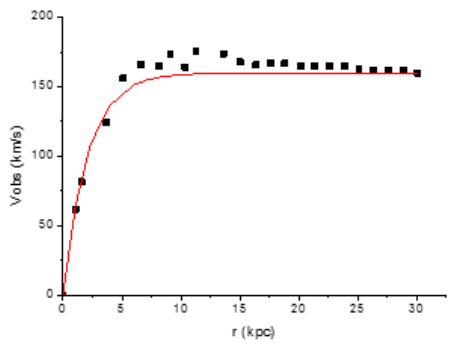
Fig. (6) NGC 1090

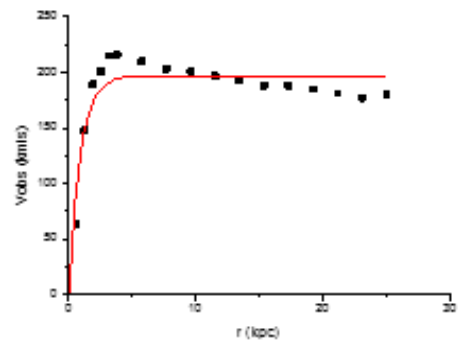
Fig. (7) NGC 2903

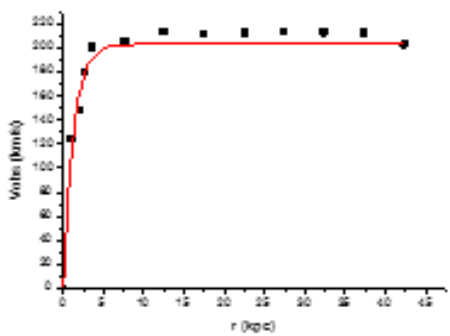
Fig. (8) NGC 2998

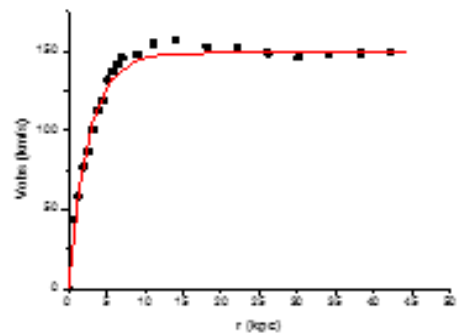
Fig. (9) NGC 3198

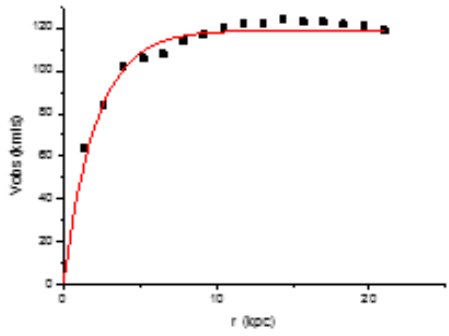

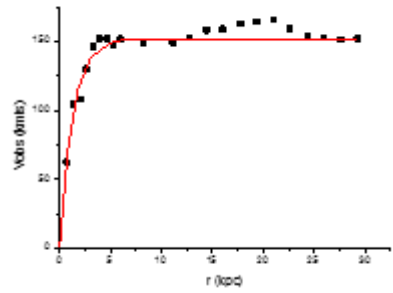

Fig. (10) NGC 4559          Fig. (11) NGC 6015

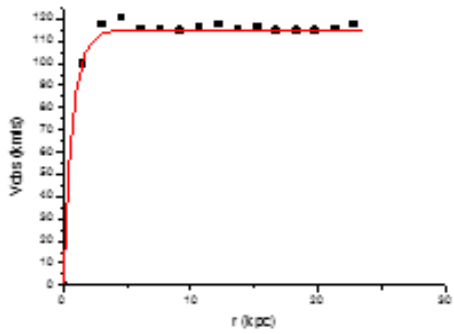

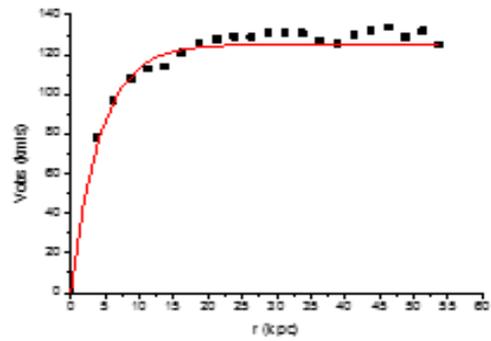

Fig. (12) NGC 6503          Fig. (13) UGC 00128

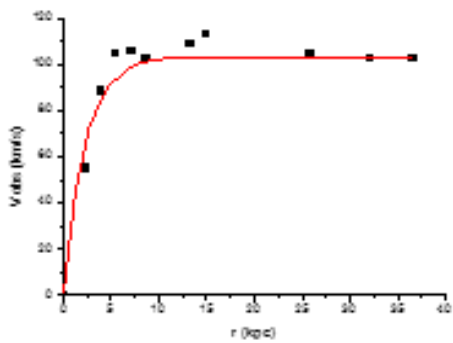

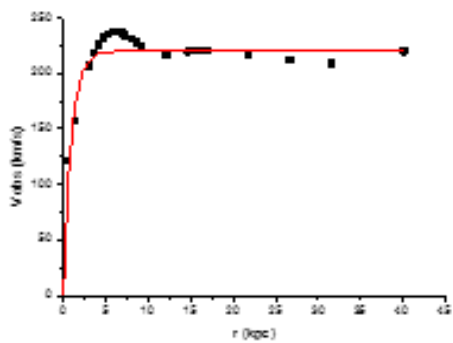

Fig. (14) UGC 01230          Fig. (15) UGC 03205

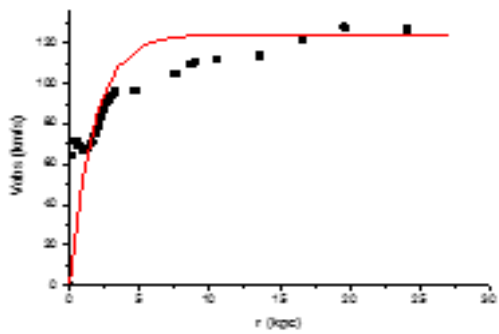

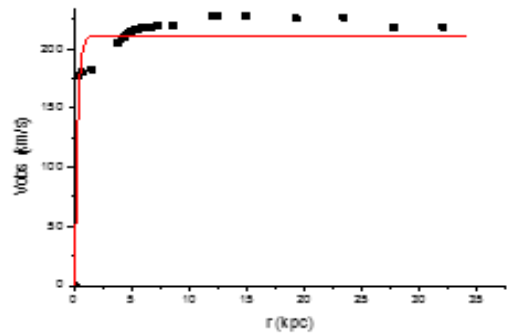

Fig. (16) UGC 03580          Fig. (17) UGC 06786

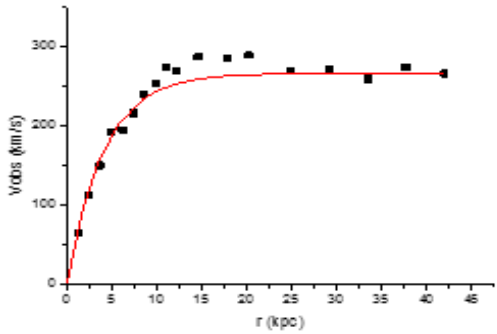
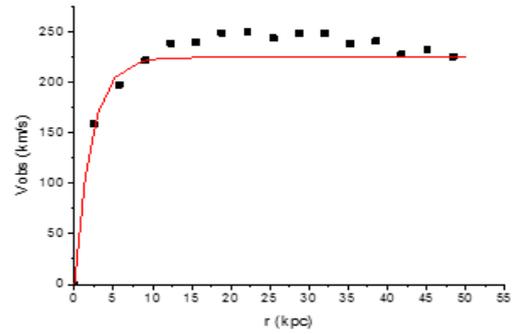

Fig. (18) UGC 11455          Fig. (19) UGC 12506

## Discussion and Conclusions

In this paper, we have tried to test the idea that the rotation curves of spiral galaxies could be resulting from its motion of its parts in a viscous medium. No attempt in this work is made to show how such viscosity is generated, the aim is only to test the assumption that once the stars are assumed to be moving in a viscous medium then the rotation curves of the galaxies can be explained accordingly as a dynamic effect.

As for the viscosity function, and since the viscosity of the medium is thought to emerging through the interaction of baryonic matter with virtual states of the vacuum, it is quite reasonable to expect that the viscosity of the medium will not be constant all through but will be some function of position. Accordingly, we have assumed certain position dependence that satisfies the very general boundary conditions in such a case. The fitting results we have obtained shows that such an assumption is quite plausible. The results exhibited in this work shows good agreement with observational data, which provides us with confidence that the basic idea is feasible. The figures below show that most fitting curves are in excellent agreement with observational data taking into consideration the approximations adopted by our model and the inevitable observational errors. If this is accepted then an urgent need appears to develop a theory which should rigorously demonstrate how vacuum viscosity emerges out of the interaction of moving baryonic matter with the vacuum states. However, it should be noted that such a medium need not to have its viscosity as an independent intrinsic property, rather the proposed viscosity is an emergent dynamic property which becomes available as masses move through quantum vacuum. Evidently, here at this point the theory is not yet established and some serious theoretical work is needed.

It worth mentioning here that the drag force will cause the parts of the galaxy at the far rim to be moving with very low acceleration as these parts reach their terminal velocity. Here our model meets with the MOND proposal [MOND], though the reasoning provided here is more profound and may have better physical explanation in addition to the fact that it applies directly on the cosmological scale. We need not to search for arguments, motion of baryonic matter in vacuum or near vacuum exhibit some emergent viscous force that hampers the expansion of the universe.

The emergent gravity proposal suggests the presence of entropic force showing elastic effect that causes higher gravity than expected on bases of standard general relativity [24].This might be compared with the viscous force we are suggesting here which causes a drag that effectively might be compared with the extra emergent gravity. However, recent investigations shows that there are some reservations on the emergent gravity proposal in relation to the Radial Acceleration Relation does not explain rotation curves of spiral galaxies except on applying certain constrains on the mass-to-light ratio [25].

McGaugh, Lelli and Schombert [27] studied the radial acceleration traced by rotation curves and that predicted by the observed distribution of baryons in galaxies with different morphologies. They found a strong correlation clearly indicating that the dark matter contribution is fully specified by that of the baryons. Now, if we take the dark matter effect to be replaced by the emergent viscosity effect we can fairly consider this finding as a supporting evidence for the case of emergent viscosity postulated in this work since such a viscosity is thought to be generated out of an interaction between quantum vacuum states and the baryonic matter.

If the notion of viscous cosmic medium is to be adopted, then the proposal of dark matter could be replaced entirely on a cosmic scale too; the expansion of the universe is hampered by the emergent viscosity of the medium. This makes the universe expands slower than expected. A slower expansion might indicate that the universe contains more matter, as such the average density of the universe will appear higher than expected and consequently a dark matter assumption may be invoked. But as we know now that the expanding medium itself is suffering from the drag caused by the viscosity thus, in principle, we are in no need for dark matter. Certainly some detailed calculations are needed once we can establish the theory for such an emergent viscosity to calculate the actual value of Hubble's parameter and compare it with observations. This will be another test of the theory besides what has been suggested for the rotation curves of the galaxies.